\documentclass{aa}\usepackage{graphics,epsf}
\newcommand{\lta}{\;
  \raise0.3ex\hbox{$<$\kern-0.75em\raise-1.1ex\hbox{$\sim$
  }}\;\hskip-2pt }
\newcommand{\gta}{\;
  \raise0.3ex\hbox{$>$\kern-0.75em\raise-1.1ex\hbox{$\sim$
  }}\;\hskip-2pt }
\begin{document}
%______________________________________________________________________
%\thesaurus{06(06.13.1; 02.03.1; 02.13.1)}
\title{The influence of density stratification and multiple
nonlinearities on solar torsional oscillations}
%______________________________________________________________________
\author{Eurico Covas\thanks{e-mail: e.o.covas@qmul.ac.uk}\inst{1}
\and David Moss\thanks{e-mail: moss@ma.man.ac.uk}\inst{2}
\and Reza Tavakol\thanks{e-mail: r.tavakol@qmul.ac.uk}\inst{1}}
%\author{Authors}
%______________________________________________________________________
\institute{Astronomy Unit, School of Mathematical Sciences,
Queen Mary, University of London, Mile End Road, London E1 4NS, UK
\and
Department of Mathematics, The University, Manchester M13 9PL, UK
}
%______________________________________________________________________
\date{Received ~~ ; accepted ~~ }
\offprints{\em R.~Tavakol}
\titlerunning{Density stratification,
multiple nonlinearities and solar torsional oscillations}
\markboth{Authors et al.: Density stratification, 
multiple nonlinearities and solar torsional oscillations}
{Authors et al.: Density stratification, 
multiple nonlinearities and solar torsional oscillations}
%______________________________________________________________________
\abstract{
%______________________________________________________________________
Analyses of recent helioseismic
data have produced ample evidence
for substantial dynamical variation of
the differential rotation
within the solar convection zone.
Given the inevitable difficulties in
resolving the precise nature of variations at deeper layers,
much effort has recently gone into determining theoretically
the expected modes of behaviour, using 
nonlinear dynamo models. Two important limitations of these models are 
that they have so far included only one form of nonlinearity,
and as yet they have not taken into account the density 
stratification in the solar convection zone.
Here we address both of these issues by 
studying the effects of including 
density stratification, as well as including an
$\alpha$--quenching nonlinearity in addition to the previously studied effects 
of the Lorentz force on the differential rotation.
We find that observationally
important features found in the earlier
uniform density models remain qualitatively unchanged, although there are 
quantitative differences. This is important as it provides 
more realistic theoretical predictions 
to be compared with and guide observations, especially in the deeper
regions where the uncertainties in the inversions are larger. 
However the presence of an effective alpha-quenching nonlinearity 
significantly reduces the amplitudes of the oscillations.

}
\maketitle
%______________________________________________________________________
\section{Introduction}
%______________________________________________________________________

Recent analyses of the helioseismic data
have produced a wealth of information concerning
the dynamical modes of behaviour associated with the
differential rotation
in the solar convection zone (CZ).
Previous analyses of data from the
Michelson
Doppler Imager (MDI) instrument on board the SOHO
spacecraft (Howe et al.\ 2000a) and
the Global Oscillation Network Group (GONG)
project (Antia \& Basu 2000)
have found that torsional oscillations
penetrate into the convection zone,
to depths of at least 10 percent in radius,
but did not agree on the behaviour 
in the poorly resolved region
near the bottom of the convection zone.
In order to understand these observed dynamical phenomena 
theoretically, and partly motivated by these apparent discrepancies,
we have previously made detailed studies of the observed variations in 
the differential
rotation in the CZ in the framework of 
nonlinear dynamo models
which included the nonlinear action
of the azimuthal component of the Lorentz force of the
dynamo generated magnetic field on the solar
angular velocity (Covas et al.\ 2000a,b; Covas et al.\ 2001a,b;
Tavakol et al. 2002). These studies produced a 
number of important robust features, that may be treated as 
their predictions.
These include: (i) the robust existence of torsional oscillations 
that penetrated all the way down to the bottom of the convection zone
for near-critical and moderately supercritical
dynamo regimes; (ii) the existence of parameter ranges for which the supercritical
models show spatiotemporal fragmentation, with different oscillatory
modes of behaviour at the top and the bottom of the CZ; (iii)
the existence of equatorial as well as polar branches 
in the angular velocity residuals; and (iv)
the presence of secondary modulations on these equatorial branches.
Predictions (i) and (ii) are in principle able to
account for different aspects of earlier observations. 
Interestingly,
the most recent inversions by 
Vorontsov et al.\ (2002a) demonstrate further important agreement
with our predictions. In particular 
they provide strong evidence that
these oscillations do in fact penetrate
down to the bottom of the CZ.
Furthermore they find clear evidence for the presence of polar branches
as well as of modulations on the equatorial branch
(see  Vorontsov et al. 2002b for preliminary results, and also
Vorontsov et al. 2003).

Despite these successes, there are two
important issues concerning these models that 
require further study. Firstly, they ignore an important 
feature of the real Sun, namely
the presence of a strong density stratification
in the solar convection zone and, secondly,
they involve only one form of nonlinearity --
that due to the Lorentz force acting on the large-scale
velocity field (here just the rotational flow).
A more realistic model would be expected to include
other forms of nonlinearity, such as the effects of
the Lorentz force on the small-scale motions, for example
as naively  modelled by a nonlinear $\alpha$--quenching.
Investigating the consequences of these shortcomings is 
particularly important, 
given the limitations on
the length of the helioseismic data sets and the
resulting uncertainties in the  inversions, especially
regarding the behaviour in the lower parts of the CZ
(see e.g. Vorontsov et al 2002).

The aim of this paper is to improve our previous model by
(i) including
a density stratification in our
previous model,
and (ii) by adding a second form of nonlinearity
in the form of  an explicit $\alpha$--quenching.
In contrast, in our previous work we imposed a radial variation of
alpha that we felt might be consistent with a strong quenching in the overshoot
region where the magnetic field was expected to be strong
(by putting $\alpha=0$ in this region).
We can now  make a detailed study of 
both the effects of density stratification and
the competing forms of nonlinearity on various
observationally important features of our model. 
Throughout the underlying zero order angular velocity
is taken to be consistent with
the recent inversions of the helioseismic (MDI) data.

The structure of the paper is as follows.
In the next section we outline our model,
and Sects.~\ref{resden} and \ref{resalp}  contain our detailed results.
In Sect.~\ref{disc}   we attempt to draw some conclusions.

%----------------------------------------------------------------------
\section{The model}
\label{mod}
%----------------------------------------------------------------------

\subsection{Equations}
\label{eqns}

We write the large-scale velocity field in the form
$\vec{u}=v\hat\phi-\frac{1}{2}\nabla\eta$, and put $v=v_0+v'$,
$v_0=\Omega_0 r\sin\theta$ being the underlying (assumed known) solar
rotation law. Thus we solve the standard mean-field dynamo equation
%----------------------------------------------------------------------
\begin{equation}
\frac{\partial\vec{B}}{\partial t}=\nabla\times(\vec{u}\times \vec{B}+\alpha\vec{
B}-\eta\nabla\times\vec{B}),
\label{mfe}
\end{equation}
%----------------------------------------------------------------------
for the mean magnetic field and the corresponding equation

\begin{equation}
\frac{\partial v'}{\partial t}=\frac{(\nabla\times\vec{B})\times\vec{B}}{4\pi\rho
r \sin\theta} . \mathbf{\hat \phi}  + \rho^{-1}\nabla.\mathbf{\tau},
\label{NS}
\end{equation}
%----------------------------------------------------------------------
($\mathbf{\tau}$ the stress tensor)
for the deviations from the underlying ($\sim$ mean) rotational 
velocity $v_0$;
as usual $\eta$ and $\nu$ are turbulent
transport coefficients.
We proceed as described in Moss \& Brooke (2000) and in previous papers in
this series, 
Nondimensionalizing in terms of the solar radius $R$ and a  time $R^2/\eta_0$,
where $\eta_0$ is the maximum value of $\eta$, and
putting $\Omega=\Omega^*\tilde\Omega$, $\alpha=\alpha_0\tilde\alpha$,
$\eta=\eta_0\tilde\eta$, $\vec{B}=B_0\vec{\tilde B}$ 
and $v'= \Omega^* R\tilde v'$.
This gives
dynamo parameters $R_\alpha=\alpha_0R/\eta_0$, $R_\omega=\Omega^*R^2/\eta_0$,
$P_{\rm r}=\nu_0/\eta_0$,  where $\Omega^*$
is the solar surface equatorial angular velocity and $B_0=(\Lambda 4\pi\rho_0\eta_0\Omega^*)^{1/2}$, 
where $\Lambda$ is an arbitrary scaling factor and 
$\rho_0$ is the density at the lower boundary of the model.
As discussed in Moss \& Brooke (2000), the physical field $B_0\vec{\tilde B}$
is independent of the value of $\Lambda$ -- for `historical' reasons we set
$\Lambda=10^4$.
$P_{\rm r}$ is the turbulent Prandtl number.

The model includes a density stratification
discussed in Section \ref{strat} below, as well as a second
source of nonlinearity in the form of an  $\alpha$--quenching
discussed in Section \ref{quench} 
below.

In this investigation, $\Omega_0$ is given in
$0.64\leq r \leq 1$ by an interpolation on the MDI data
obtained from 1996 to 1999 (Howe et al. 2000a).
For $\alpha$ we put $\alpha=\alpha_{\rm r}(r)f(\theta)$,
where $f(\theta)$ was chosen to be $\sin^2\theta\cos\theta$.
% or $\sin^4\theta\cos\theta$.
This form has been used previously (see e.g. R\"udiger \& Brandenburg 1995)
and its choice here is simply
to make the butterfly diagrams more realistic.
For $\alpha_{\rm r}$ we took the following choices.
In the absence of $\alpha$--quenching,
$\alpha_{\rm r}$ was set to be 
$\alpha_{\rm r}=1$ for $0.7 \leq r \leq 0.8$
with cubic interpolation to zero at $r=r_0$ and $r=1$.
In presence of $\alpha$--quenching,
we took $\alpha_{\rm r}=1$ throughout the 
convection zone.
%$\alpha_{\rm r}=1$ in part or all
%of the CZ (see below for details),
%with cubic interpolation to zero at $r=r_0$ and $r=1$,
We use the convention that $\alpha_{\rm r}>0$
and $R_\alpha < 0$, to get the correct sense of migration of field patterns. 
Also, in
order to take into token account the
likely decrease in the turbulent diffusion coefficient $\eta$
in the overshoot region, we allowed a simple
linear decrease from $\tilde\eta=1$ at $r=0.8$
to $\tilde\eta=0.5$ in $r<0.7$.

At the surface $r=R$ we
took vacuum boundary conditions, whereby
the poloidal field within $r=R$ is
smoothly joined, by a matrix multiplication, 
to an external vacuum solution;
the azimuthal field there satisfies  $B=0$. 
At the inner boundary $r=r_0$ we used the same conditions as
Tavakol et al. (2002). We note that `reasonable' changes in these
boundary conditions
do not qualitatively change our conclusions 
(cf. Tavakol et al. 2002).

We set  $r_0=0.64$; with
the solar CZ proper being thought to occupy the region $r \gta 0.7$,
the region $r_0 \leq r \lta 0.7$ can be thought of as an overshoot
region/tachocline, where the major radial gradients of $\Omega$ occur.
Equations (\ref{mfe}) and (\ref{NS}) were solved using a
modified version of the code
described in Moss \& Brooke (2000), using the  
the above boundary conditions,
over the range $r_0\leq r\leq1$, $0\leq\theta\leq \pi$.
We used a mesh resolution of $61 \times 101$
points, uniformly distributed in radius and latitude respectively,
i.e. $\Delta r=0.06, \Delta\theta=0.01\pi$.

%______________________________________________________________________
\subsection{Density stratification}
\label{strat}
%______________________________________________________________________

\begin{figure}
\centerline{\def\epsfsize#1#2{0.5#1}\epsffile{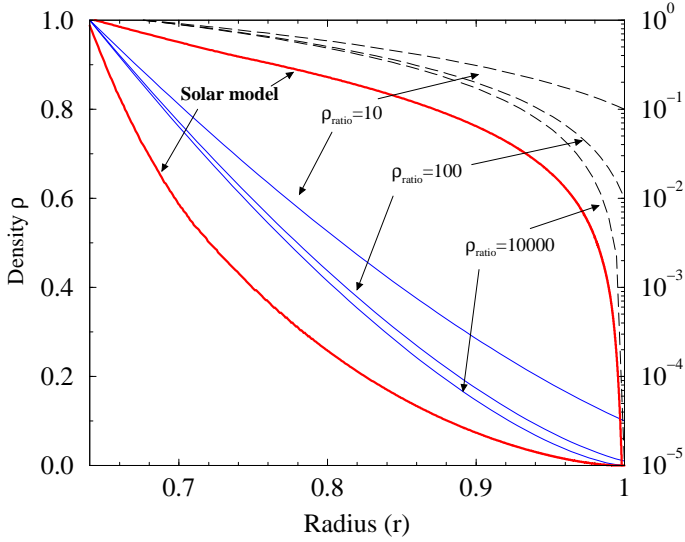}}
\caption[] {\label{0046fig1.eps}
Density profiles obtained using a
polytropic model, for several values of $\rho_{\rm ratio}$.
Here $\rho$ is dimensionless, measured in units of the density at $r=r_0$.
The upper set of curves show $\log \rho$, the lower show $\rho$.
Shown also is the density profile of CD96.
For $\rho_{\rm ratio}\ga 10^2$,
the density profile is approximately invariant through  most of convection zone, apart from the regions near the surface.
}
\end{figure}

The density profile in the solar model computed by  Christensen-Dalsgaard et al.
(1996; henceforth referred to as CD96)
is shown in Fig.~\ref{0046fig1.eps}.
We approximated the density stratification in the solar 
convection zone by a polytropic density 
profile, writing in general
\begin{equation}
\label{density}
\rho(r)/\rho_0=\left(\frac{r_s-r}{r_s-r_0}\right)^n,
\end{equation}
where
\begin{equation}
r_s=\left(\frac{\rho_{\rm ratio}^{1/n}-r_0}{\rho_{\rm ratio}^{1/n}-1}\right);
\end{equation}
we take $n=1.5$ below.
This density profile corresponds to a polytrope truncated at  the appropriate
radius to give the required value of $\rho_{\rm ratio}$, with this radius subsequently
renormalized to unity. $\rho_{\rm ratio}$ is thus the ratio of the densities
at the top ($r=1$) and the bottom ($r=r_0$) of the computational domain; 
$\rho_0=\rho(r_0)$.
Several examples of such profiles,
with different values of $\rho_{\rm ratio}$
are also shown in Fig.~\ref{0046fig1.eps}.
For $\rho_{\rm ratio}\ga 100$, these give a reasonable approximation 
to the density stratification in
the bulk of the solar CZ, apart from the outermost regions near the
solar surface

%____________________________________________________________________
%_____________________________________________________________________

%-----------------------------------------------------
\subsection{Explicit $\alpha$--quenching}
\label{quench}
%------------------------------------------------------ 

The nonlinearity in our original model
is through the
nonlinearity in the 
Lorentz force and the subsequent presence of the  $v'$ term in
the induction equation. In  all of our previous studies
the form of
$\alpha_{\rm r}$ was chosen to be independent of the magnetic
field $\vec{B}$. It did, however, include 
an implicit $\alpha$-quenching
in the overshoot layer in that $\alpha$ was
prescribed to vanish in $r\leq 0.7$
and to be relatively small in the layers immediately above where
$|\vec{B}|$ was anticipated to be relatively large.
%Otherwise,
%it was assumed to be independent of the magnetic
%field $\vec{B}$.
Here we study the effects of having an 
additional nonlinearity in the form of an
explicit nonlinear $\alpha$-quenching 
given (reverting temporarily to dimensional units) by

\begin{equation}
\label{quenching-proper}
\alpha= \alpha_{\rm r}(r) \frac{\sin^2\theta\cos\theta}{1+  g' |\vec{B}^2|/B_{\rm eq}^2}
\end{equation}
where 
$B_{\rm eq}=(4\pi\rho v_t^2)^{1/2}$ is the equipartition field strength,
${v_t}$ is the turbulent velocity
and $g'$ is of order unity.    
(The trigonometric factor is chosen to agree with that used in  our earlier 
work -- see above.)
In order to estimate how $B_{\rm eq}^2$, appearing in 
the denominator, varies with depth
we recall that to a very good approximation the 
luminosity $L= 4\pi r^2 F_r \propto  r^2 \rho {v_t}^3$ is a constant in the CZ,
which gives $v_t \propto \rho^{-1/3} r^{-2/3}$ and hence
$\rho {v_t}^2 \propto \rho^{1/3} r^{-4/3}$. This suggests that
the denominator in (\ref{quenching-proper}) will be 
a slowly varying function of the depth,
especially away from the solar surface, and this
expectation is supported by detailed mixing length models of the solar CZ.
Thus we take $B_{\rm eq}^2\propto \rho v_t^2={\rm const}$,  and so
the $\alpha$-quenching expression (\ref{quenching-proper}) can be well
approximated (now temporarily using dimensionless variables again) by

\begin{equation}
\label{quenching}
\tilde\alpha= \alpha_{\rm r}(r) \frac{\sin^2\theta\cos\theta}{1+ g|\vec{B}^2|},
\end{equation}
where $g=g'\Lambda4\pi\rho_0\eta_0\Omega^*/B_{\rm eq}^2$. $B_{\rm eq}^2$ is constant,
and thus so is $g$.
(In this connection it may be noted that there is an ongoing controversy
regarding the nature and strength of $\alpha$-quenching.
This is not the place to rehearse the arguments for and against `strong'
alpha-quenching (where $g'\gg 1$); the issue is unresolved, and 
we use (\ref{quenching}) as a commonly adopted nonlinearity.)

We can make an estimate of what might be plausible values for $g$ by taking
$\eta_0=3\times 10^{11}$ cm$^2$sec$^{-1}$n (see Sect.~\ref{resden} below), 
$\Lambda=10^4$, $\rho_0=1$ gm cm$^{-3}$,
$\Omega^*=2\times 10^{-6}$ sec$^{-1}$, $B_{\rm eq}= 3500$ G 
(the latter estimated from an envelope model by Baker \& Temesvary (1966) (more
recent models do not give significantly different values).
Then $g\sim 3\times 10^3 g'$. (As noted above, $B_0^2, g \propto \Lambda$, but
$\vec{\tilde B}\propto \Lambda^{-1/2}$, so the results in physical variables 
do not depend on the value adopted for $\Lambda$.)

The inclusion of both types of 
nonlinearity in the model allows us to study the 
relative importance of each term and in particular
what happens to the amplitudes of the 
torsional oscillations as
$g$ increases.

%_______________________
\section{Results: effects of density stratification}
\label{resden}
%_______________________
Using the above model, we studied the dynamics in the convection zone,
and in particular the changes caused by the introduction
of the density stratification 
and the additional nonlinearity. 
We concentrated on the effects of these
changes on the torsional oscillations
and in particular on their amplitudes
(i.e. the angular velocity residuals,
defined as  the current  angular velocity 
minus the background angular velocity $\Omega_0$),
since these are the most important
indicators to be compared with observations.
The model predictions are likely to be of particular importance
in deeper regions where
the uncertainties in the inversions are largest.

We first  study the effects of density stratification in isolation,
with $g=0$. We fixed $R_\omega=6\times 10^4$ (corresponding to $\eta_0\approx 3\times 10^{11}$ cm$^2$sec$^{-1}$) by requiring that the dynamo
period is near to 22 years close to marginal stability. The marginal value of
$R_\alpha$ is then about $-3.2$.
%______________________________________________
\subsection{Comparison with previous results}
%______________________________________________
%______________________________________________________________________
\begin{figure}
\centerline{\def\epsfsize#1#2{0.43#1}\epsffile{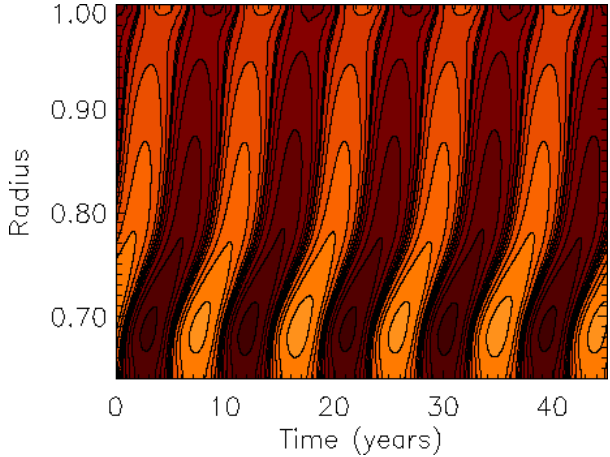}}
\caption[]{\label{0046fig2.eps}
The radial (r--t) contours of the angular velocity
residuals $\delta \Omega$ as a
function of time for a cut at $10$ degrees latitude.
The model parameters are
$R_\alpha = -4.0, P_{\rm r} = 1.0, R_\omega =60000$
and $\rho_{\rm ratio} =10000$. 
Note the penetration of the torsional oscillations all the way down to the bottom
of the convection zone.
Darker and lighter regions represent positive and negative
deviations from the time averaged background rotation rate.
}
\end{figure}
%______________________________________________________________________

To begin with we made a detailed comparison of 
the results obtained here in the presence of  density stratification,
with our previous studies which ignored this 
effect.
In particular,
we concentrated on one of the findings of these earlier studies,
namely the fact that  
torsional oscillations were found to extend all the way
down to the bottom of the CZ.

Our results here show  that, for both near-critical, 
and moderately supercritical
dynamo regimes, the oscillations extend all
the way down to the bottom of the
convection zone, for all $1\le \rho_{\rm ratio} \le 10^6$.
% in the range $ \rho_{\rm ratio} \in [1, 10^6]$. 
This is in line with  our previous results which shows
that such behaviour is an extremely robust feature
of the models considered to density stratification.
An example of such generic penetration
is depicted in Fig.~\ref{0046fig2.eps}.
%Such penetration is an extremely robust feature
%of the models considered, and is qualitatively unchanged by the inclusion of
%density stratification. 
This is important observationally since
the inversions have so far not been able to resolve unambiguously
regimes near the bottom of the convection zone (Howe et al.\ 2000a; Antia \& Basu 2000).

\begin{figure}
\centerline{\def\epsfsize#1#2{0.43#1}\epsffile{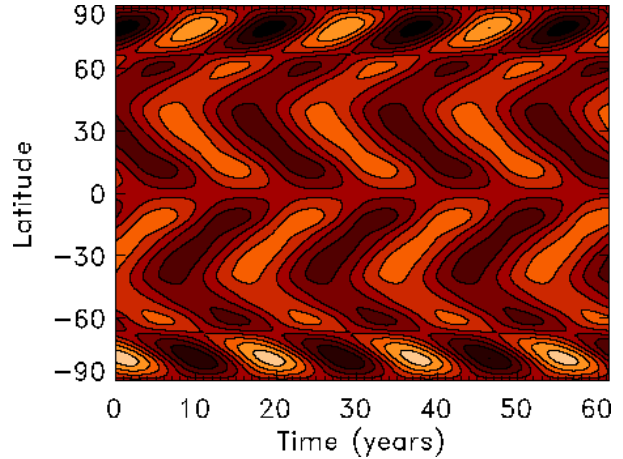}}
\caption[]{\label{0046fig3.eps}
Butterfly diagram of the toroidal component of the
magnetic field $\vec{B}$ at
$r=0.99$. The 
parameters are the same as those 
for Fig.~\ref{0046fig2.eps}.
Dark and light shades correspond to positive and negative values of
$B_\phi$ respectively.
}
\end{figure}

\begin{figure}
\centerline{\def\epsfsize#1#2{0.43#1}\epsffile{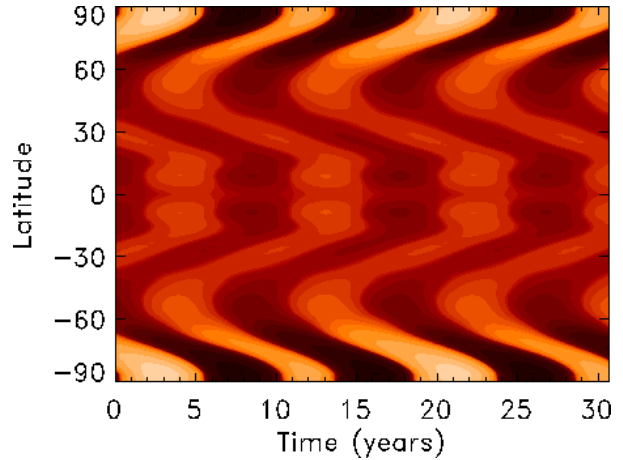}}
\caption[]{\label{0046fig4.eps}
Angular velocity residuals at $r=0.99$ with latitude and time.
The
parameters are the same as those
for Figure \ref{0046fig2.eps}.
A temporal average has been subtracted to reveal the migrating banded zonal
flows. Darker and lighter regions represent positive and negative
deviations from the time averaged background rotation rate.
}
\end{figure}

For completeness we also show 
in Fig.~\ref{0046fig3.eps}
the butterfly diagram (i.e. latitude-time plot) for the toroidal component 
of the magnetic field $\vec{B}$ near the surface,
as well as the corresponding plot for the
%Fig.\ \ref{ivac=7.velocity.C=0.92.eps} shows the
angular velocity residuals 
%at $R=0.92$
(i.e. the near-surface torsional oscillations)
in Fig.~\ref{0046fig4.eps}.
As can be seen, the presence of significant density
stratification does not alter the qualitative behaviour of the
magnetic field and torsional oscillations.
%______________________________________________________________________
%______________________________________________________________________

%______________________________________________________________________
%______________________________________________________________________

%______________________________________________________________________
\subsection{Amplitudes of oscillations as a function of $\rho_{\rm ratio}$}
%______________________________________________________________________

$\quad$ From an observational point of view,
an important feature of the  dynamics of the CZ is
the way the amplitudes of the 
torsional oscillations vary as a function of
model ingredients and parameters, as well as depth
in the CZ. A theoretical understanding is 
crucially important here, especially given the uncertainties 
in inversions of the helioseismic data in 
the lower parts of the CZ.

\begin{figure}
\centerline{\def\epsfsize#1#2{0.43#1}\epsffile{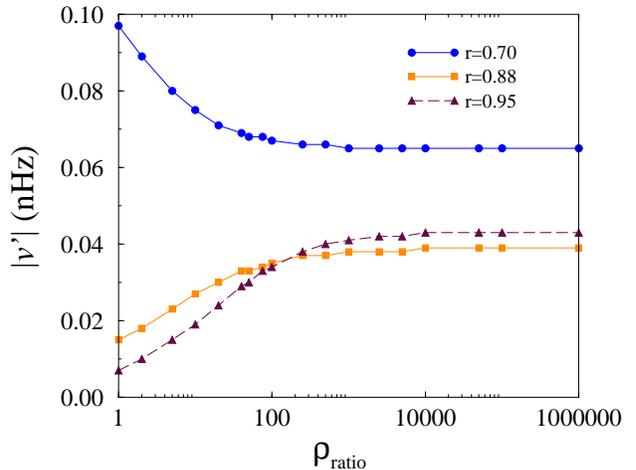}}
\caption[] {\label{0046fig5.eps}
The variation of oscillation amplitudes at $r=0.70, 0.88$ and $0.95$
as a function of
$\rho_{\rm ratio}$, for a near-onset dynamo regime. The parameters are 
$R_\alpha =-3.2, P_{\rm r} =1.0, R_\omega=60000$, with $\alpha=\alpha_r(r)f(\theta)$
and no $\alpha$-quenching.
}
\end{figure}

\begin{figure}
\centerline{\def\epsfsize#1#2{0.43#1}\epsffile{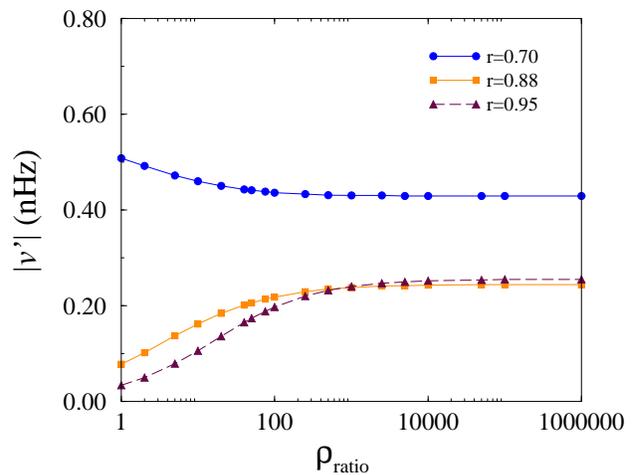}}
\caption[] {\label{0046fig6.eps}
The variation of oscillation amplitudes at $r=0.70, 0.88$ and $0.95$
as a function of
$\rho_{\rm ratio}$, for a supercritical case. The parameters are        
$R_\alpha =-4.4, P_{\rm r} =1.0, R_\omega=60000$, with $\alpha=\alpha_r(r)f(\theta)$
and no $\alpha$-quenching.
}
\end{figure}

With this in mind, we studied how the amplitudes of the torsional oscillations
change as a function of
$\rho_{\rm ratio}$, for two values of $R_\alpha$;
one corresponding to a near-critical and the other
to a supercritical dynamo regime.
The results are depicted in Figs.~\ref{0046fig5.eps}
and \ref{0046fig6.eps}.
For orientation, we recall that the corresponding amplitudes
in the sun, deduced from helioseismological observations, are of the order of one nHz.

The important trend revealed by these figures is that,
in both the near--critical and supercritical cases,
the amplitudes of the torsional oscillations
in the upper part of the CZ increase as
$\rho_{\rm ratio}$ increases, while those at the bottom
decrease. 
The decrease in amplitude of the
torsional oscillations in the bottom of
the convection zone, as $\rho_{\rm ratio}$ increases
(i.e. as the density profile becomes more realistic)
is of practical importance, as it suggests that
the detection of such oscillations may be even more difficult
than might appear at present just from the
limits to the resolution of the helioseismological
inversions at the bottom of the convection zone.
The increase of the amplitudes at the top is also
important since it improves the agreement between simulated and observed values.

\begin{figure}
\centerline{\def\epsfsize#1#2{0.45#1}\epsffile{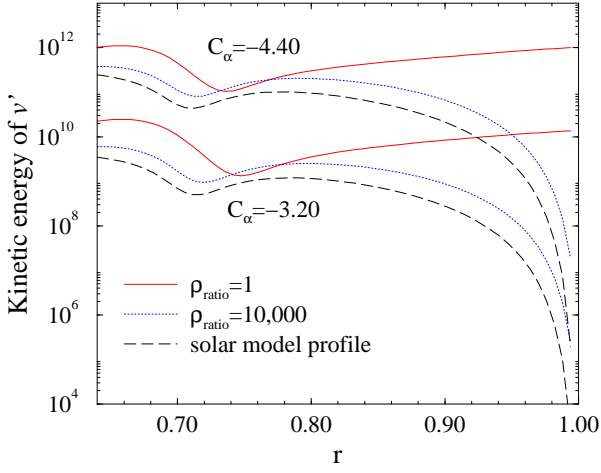}}
%\centerline{\def\epsfsize#1#2{0.43#1}\epsffile{0046fig7.eps}
\caption[] {\label{0046fig7.eps}
The distribution of the kinetic energy (in arbitrary units) in spherical shells 
of thickness the radial mesh size $\Delta r$,
as a function of radius and $\rho_{\rm ratio}$, for both near-onset
and supercritical dynamo regimes. 
The parameters are as in Figs.\ \ref{0046fig5.eps}
and \ref{0046fig6.eps} respectively.}

\end{figure}

Also of potential  interest is the way that the distribution of 
the perturbation kinetic energy,
defined by $\int_V  \rho(2v_0v'+v'^2) dV$, is
modified as the density stratification changes.  
In Fig.~\ref{0046fig7.eps} we show the variation of this integral with
radius, where we take the domain $V$ of integration to be the local spherical 
shell with thickness $\Delta r$.
Introducing a stratification markedly reduces the perturbation kinetic energy
in the near-surface regions, but the differences between the two
stratified models illustrated are relatively small. However the low values
of the near-surface density in the stratified models means that the amplitude
of the torsional oscillations in this region are {\it not} correspondingly
decreased -- see Figs.~5 and 6. This relative insensitivity of the observed
quantity to the stratification is reassuring.

%____________________________________________________________________
%_____________________________________________________________________
%%____________________________________________________________________

%%____________________________________________________________________
%%____________________________________________________________________

%______________________________________________________________________
\subsection{Amplitudes of oscillations as a function of $R_\alpha$}
%______________________________________________________________________

\begin{figure}
\centerline{\def\epsfsize#1#2{0.43#1}\epsffile{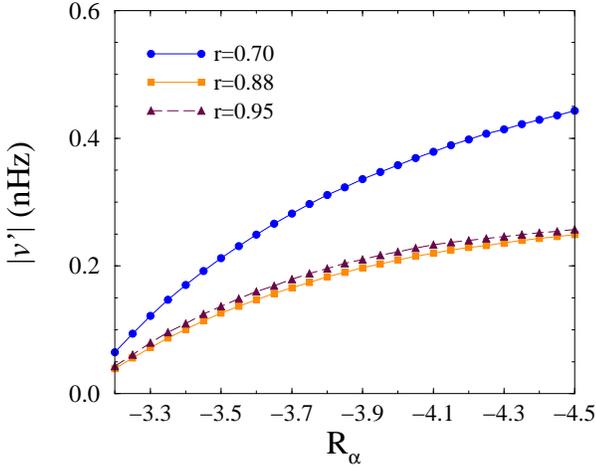}}
\caption[] {\label{0046fig8.eps}
The variation of oscillation amplitudes as a function of
$R_\alpha$. Parameter values are $P_{\rm r}=1.0$ and $R_\omega=60000$,
$\rho_{\rm ratio}=10000$, with $\alpha=\alpha_r(r)f(\theta)$ and no $\alpha$-quenching.
}
\end{figure}

\begin{figure}
\centerline{\def\epsfsize#1#2{0.43#1}\epsffile{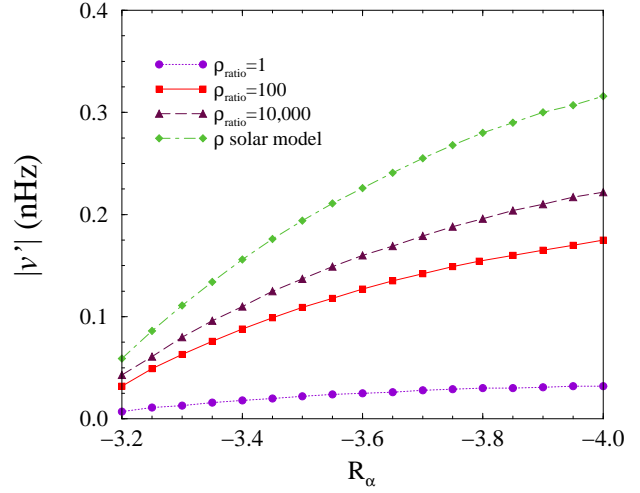}}
\caption[]
{\label{0046fig9.eps}
The variation of oscillation amplitudes as a function of
$R_\alpha$ for different values of $\rho_{\rm ratio}$ at $r=0.95$. Shown also
are the variations obtained using the solar model density profile of CD96.
The parameters are $R_\omega=60000$, $P_{\rm r} =1.0$ with $\alpha=\alpha_r(r)f(\theta)$
and no $\alpha$-quenching.
}
\end{figure}

To begin with, we verified that for
a given  density profile (in presence of density stratification), 
the amplitudes of the oscillations
increase as the dynamo number $R_\alpha$ is increased
(see also Covas et al 2001a).
The results are shown in
Fig.~\ref{0046fig8.eps}.
We then studied how the amplitudes change as
we change the density profile by increasing
the value of $\rho_{\rm ratio}$. For comparison we also
calculated the amplitudes for the  density profile of CD96.
The results are shown in
Fig.~\ref{0046fig9.eps}.
%Fig.~\ref{amplitudes.Calpha.MDI.rhorats.eps}.
As can be seen, the effect of increasing
$\rho_{\rm ratio}$ is to increase the amplitudes of
oscillations near the surface. The largest 
amplitudes are attained for the CD96 density distribution.
%For comparison we have also plotted
%in Fig.~\ref{0046fig9.eps} the amplitudes near the surface
%as a function of $R_\alpha$ for various values of $\rho_{\rm ratio}$
%as well as for a model density profile.

%____________________________________________________________________
%_____________________________________________________________________

%____________________________________________________________________
%_____________________________________________________________________

%______________________________________________________________________
\subsection{Amplitudes of oscillations as a function of depth and $\rho_{\rm ratio}$}
\begin{figure}
\centerline{\def\epsfsize#1#2{0.43#1}\epsffile{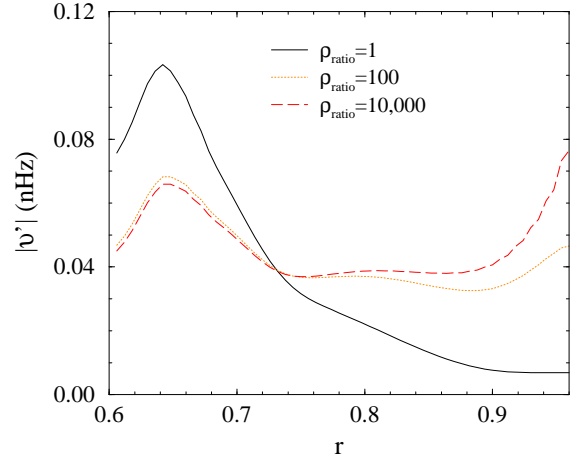}}
\caption[] {\label{0046f10.eps}
The variation of oscillation amplitudes as a function of
depth for a number of values of $\rho_{\rm ratio}$, for a near-onset regime with
$R_\alpha = -3.2$, $R_\omega=60000$, $P_{\rm r} =1.0$,
$\alpha=\alpha_r(r)f(\theta)$ and 
no $\alpha$-quenching.
}
\end{figure}

\begin{figure}
\centerline{\def\epsfsize#1#2{0.43#1}\epsffile{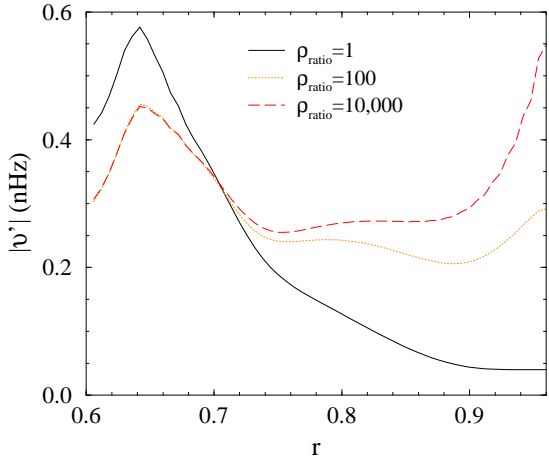}}
\caption[] {\label{0046f11.eps}
The variation of oscillation amplitudes as a function of
depth for a number of values of $\rho_{\rm ratio}$, for a supercritical case 
with
$R_\alpha = -4.4$, $R_\omega=60000$, $P_{\rm r} =1.0$,
$\alpha=\alpha_r(r)f(\theta)$ and 
no $\alpha$-quenching.
}
\end{figure}

%______________________________________________________________________
We also studied the variations of the amplitudes of oscillations as
a function of depth for a number of values of $\rho_{\rm ratio}$,
for both a near-onset regime (with $R_\alpha = -3.2$)
as well as for more supercritical parameters ($R_\alpha = -4.4$),
and the results are shown in  Figs.~\ref{0046f10.eps}
and \ref{0046f11.eps} respectively.
As can be seen, despite the latter
having larger amplitudes, the two cases have
important similarities. In each
the amplitudes near the surface layers increase with increasing
$\rho_{\rm ratio}$, whereas they decrease near the bottom.

%____________________________________________________________________
%_____________________________________________________________________

%____________________________________________________________________
%_____________________________________________________________________

%_____________________________________________________________________

\section{Results: comparative study of the nonlinearities}
\label{resalp}
%______________________________________________________________________

In order to make a comparative study of the
effects of the two competing forms of nonlinearities
considered here, we begin (see also below) by recalling that,
in the absence of any alpha-quenching, increasing the degree of 
nonlinearity due to the Lorentz force
(i.e. increasing $R_\alpha$)
%or the 
%Prantdl number
results in an increase in the amplitudes of the  torsional
oscillations. The question is
what is the effect of increasing the 
strength of the additional nonlinearity
due to $\alpha$--quenching
on these amplitudes?

\begin{figure}
\centerline{\def\epsfsize#1#2{0.43#1}\epsffile{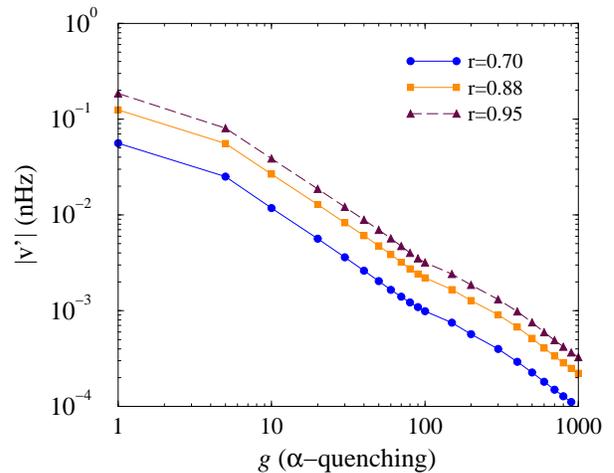}}
\caption[] {\label{0046f12.eps}
The variation of amplitudes of torsional oscillations as a function of
changes in the $\alpha$-quenching coefficient $g$ for a near-onset 
regime. Parameters are $R_\alpha =-2.5$, $P_{\rm r} =1.0$, $R_\omega=60000$
with $\alpha_{\rm r} =1$ (i.e.\ no explicit radial dependence of $\alpha$)
and the density $\rho$ given by the solar model of CD96.
}
\end{figure}

We studied this question by making a detailed study of 
how the amplitudes of torsional 
oscillations change as a function of
$g$,  the coefficient of  the
$\alpha$--quenching term (Eq.~(\ref{quenching-proper})). 
The results are presented in 
Fig.~\ref{0046f12.eps}.
A number of trends are apparent.
The amplitudes of the torsional oscillations
decrease at all depths with increasing $\alpha$-quenching coefficent $g$.
However,  the oscillations at
the top of the CZ have the highest amplitudes throughout.

\begin{figure}
\centerline{\def\epsfsize#1#2{0.43#1}\epsffile{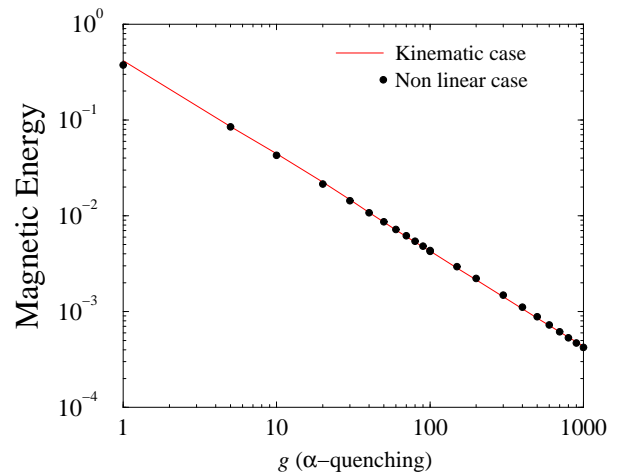}}
\caption[] {\label{0046f13.eps}
The variation of the  magnetic energy as a function of the
$\alpha$-quenching coefficient $g$. Shown are both the kinematic (without the
Navier-Stokes equation)
and the nonlinear (with the Navier-Stokes equation) cases.
Parameters are $R_\alpha =-2.5$, $P_{\rm r} =1.0$ and $R_\omega=60000$,
with $\alpha_{\rm r} =1$ 
%i.e.\ no explicit radial dependence of $\alpha$,
and the density $\rho$ given by the solar model of CD96.
}
\end{figure}

To study the influence of increasing the $\alpha$--quenching coefficient,
we recall that  in the case of pure alpha-quenching,
\begin{equation} 
\label{scaling}
<{\vec{B}}^2> \propto \frac{1}{g},
\end{equation}
where the angled parentheses denote some sort of global average.
We checked that this was 
exactly satisfied for our code in the 
quasi-kinematic approximation when the Navier-Stokes equation
was switched off. Interestingly, we also found
similar behaviour when the additional nonlinear feedback via
the Navier-Stokes equation  was present. This is shown in
Fig.~\ref{0046f13.eps}, which displays the behaviour
of the global magnetic energy as a function
of $g$.

The amplitudes of the torsional 
oscillations increase somewhat with $|R_\alpha|$
but in the range accessible to our code ($|R_\alpha|\la 15$) this increase
is insufficient to compensate for the decrease caused
by the increase in the $\alpha$-quenching coefficient $g$
(given by (7)).

We can add here that 
in our previous, very preliminary, study of
the effects of
alpha-quenching on the torsional
oscillations (Covas et al 2001a), we were only able to
find oscillatory solutions for very small values of
the alpha-quenching coefficient ($g <0.1$).
What has enabled us to find the existence of oscillatory
solutions at such large values of $g$ here
is our recent discovery of the 
existence of multiple attractors in the dynamo system,
including both oscillatory and steady solutions, 
that coexist to $g \sim 1000$, i.e. in the physically plausible range as 
estimated in Sect.~2.3.

%____________________________________________________________________
%_____________________________________________________________________
%____________________________________________________________________
%_____________________________________________________________________

%_______________________
\section{Discussion}
\label{disc}
%_______________________

We have made a detailed study of the effects 
of including 
density stratification in the model, as well as an additional form of 
nonlinearity due to $\alpha$--quenching on the behaviour
of solar torsional oscillations. We feel that the previous omission of any
discussion of these effects reduced the detailed applicability of our results
to the observational data.

To study the effects of density stratification,
first of all we made a comparison with our previous results and
found both  important similarities and differences.
Concerning similarities, we found that consistent with our previous findings
for near-critical and moderately supercritical
dynamo regimes, the torsional oscillations extend all
the way down to the bottom of the
convection zone. This is true in presence of density stratification
with all values of $\rho_{\rm ratio}$ in
the range $[1, 10^6]$.
Concerning differences, we found that
in both the near--critical and supercritical cases,
the amplitudes of the torsional oscillations
in the upper part of the CZ increase with
$\rho_{\rm ratio}$, while those at the bottom
decrease.  
Given that the uncertainties
in the inversions are expected to be greatest at
the bottom of the CZ, the lowering of amplitudes
in these regions {with increasing density stratification
has potential importance  for the analysis  
of helioseismological data, and     
for the understanding of the dynamo regimes  
near the bottom of the convection zone}.

When two forms of
nonlinearity are simultaneously present, we find that torsional oscillations persist
in presence of both nonlinearities, for 
a large range of  values of the $\alpha$--quenching coefficient $g$ up
to $g>1000$. The amplitudes of these oscillations decrease however,
as they shadow the square of the magnetic field strength,
which in turn follows the 
scaling with $g$ given by (\ref{scaling}).

We also note that for small $g$, the
combined effect of increasing $\rho_{\rm ratio}$ and $g$ is 
to leave the amplitude of the near-surface torsional oscillations little changed.
However, for larger $g$ ($\ga 10$), the effect of $\alpha$-quenching dominates,
as the amplitudes of the torsional oscillations scale as $1/g$.

Equation ~(\ref{NS}) suggests, and our models confirm,
that variations in
$v'$ closely shadow those in the
$\phi$-component of the Lorentz force.
It seems {\em a priori} difficult to imagine a driver for torsional
oscillations that does not involve the Lorentz force, either at a macro- or
micro-level (e.g. the `nonlinear $\Lambda$--effect',
Kitchatinov \& Pipin 1998;
Kitchatinov et al.\ 1999), whatever the
prime nonlinearity in the dynamo. Thus the requirement that the Lorentz force
shadows the torsional oscillations is plausibly an additional constraint
even on dynamo models that do not explicitly include dynamical effects
(e.g. conventional $\alpha$-quenched models. Brandenburg \& Tuominen 1988
appear to have been the first to investigate semi-quantitatively such  an idea.)
However it is perhaps a little puzzling that values of $g$ as large 
as suggested by the estimates of Sect.~\ref{quench}
reduce the amplitudes of the torsional oscillations so substantially,
removing the previous agreement with observations.

%______________________________________________________________________

%______________________________________________________________________
\end{document}